\documentclass[preprint,showpacs,aps,prl]{revtex4-1}

\usepackage{amsmath}
\usepackage{amssymb}
\usepackage{graphicx}
\usepackage{bbold}

\newcommand{\bvec}{\mathbf}
\begin{document}

\title{Ultrafast Spin-Lattice Relaxation in Ferromagnets Including Effective Spin-Orbit Fields}
\author{Kai Leckron}
\author{Svenja Vollmar}
\thanks{Also with Graduate School of Excellence \emph{Materials Science in Mainz}, 67663 Kaiserslautern, Germany}

\author{Hans Christian Schneider}

\email{hcsch@physik.uni-kl.de}
\affiliation{Physics Department and Research Center OPTIMAS, Kaiserslautern University, P. O. Box 3049, 67663 Kaiserslautern, Germany}

\date{\today}

\begin{abstract}
We investigate ultrafast demagnetization due to electron-phonon interaction in a model band-ferromagnet. We show that the microscopic mechanism behind the spin dynamics due to electron-phonon interaction is the interplay of scattering and the precession around momentum-dependent effective internal spin-orbit magnetic fields. The resulting magnetization dynamics can \emph{only} be mimicked by spin-flip transitions \emph{if} the spin precession around the internal fields is sufficiently fast (compared to the scattering time) so that it averages out the transverse spin components. 
\end{abstract}

\pacs{XXXXX}

\maketitle
\section{Introduction} 

In 3d-ferromagnets, excitation by an ultrashort linearly polarized pulse can reduce the magnetization, as observed by the magneto-optical Kerr effect~\cite{Beaurepaire.1996,Krauss:2009gc} or X-ray magnetic circular dichroism~\cite{Stamm:2007hy}, by 50\% and more, even reaching a complete ``quenching'' of the magnetization for high fluence. The spin angular momentum, as determined experimentally, is thus dramatically reduced. Experimental evidence~\cite{Eschenlohr:2013id,Vodungbo:2016gt} points mainly to the importance of electronic scattering and transport for this effect.

 As the transport contribution to magnetization can be suppressed, there must be an additional microscopic mechanism that contributes to the observed magnetization dynamics, which is still under debate~\cite{ZhangHuebner:2000prl,Carva:2011nature-comment,Krieger:2015io,Tows:2015jk}. Here we focus on the mechanism that has long been regarded as the most probable explanation of demagnetization dynamics: electronic spin-flip scattering with phonons, which is often called the Elliott-Yafet demagnetization mechanism~\cite{Koopmans:PRL2005,Steiauf.2009} after a spin relaxation mechanism for electrons in semiconductors~\cite{Yafet:1963,Elliott:1954,Overhauser:1953,Baral:2016cr}. 
 
The original Elliott-Yafet mechanism was developed for a pair of \emph{degenerate bands}, whose non-pure spin states are of the general form  $|\bvec{k}, \pm\rangle= a^{(\pm)}_{\bvec{k}}|\!\!\uparrow\rangle + b^{(\pm)}_{\bvec{k}}|\!\!\downarrow\rangle$ due to spin-orbit coupling. The labels ``$+$'' and ``$-$'' indicate whether a state is predominantly spin-up or spin down, depending on which coefficient $|a|^2$ or $|b|^2$ is larger. 
 In ferromagnets, the majority and minority states are of the same general form, but due to the spin splitting the labels ``$+$'' and ``$-$'' now also refer to the spin eigenvalues with respect to a quantization axis. Due to spin-orbit coupling this quantization axis is $k$ dependent and the spin structure belonging to the $+$/$-$ states is essentially noncollinear. The conventional Elliott-Yafet mechanism determines the spin dynamics due to electron-phonon interactions from transition rates between $+$ and $-$ states and is therefore incapable of describing deviations from the spin quantization axis, i.e., spin coherences. However, these coherences are always present and may be expected to be particularly important if the noncollinearity is prounounced, for instance, at spin-orbit hybridization points in band ferromagnets~\cite{Weinelt:PRL08}. 

 In the present paper, we include \emph{consistently} the influence of spin-orbit coupling and exchange splitting on electron-phonon scattering dynamics and numerically study the case of a model ferromagnet. We obtain the ensemble magnetization dynamics from the microscopic spin-density matrix, which allows us to include spin coherences in noncollinear (i.e., non-trivially $k$-dependent) spin structures. While we do not present a complete theory of demagnetization dynamics, we demonstrate how spin-flip transitions, as they are assumed in the conventional Elliott-Yafet mechanism and believed to play an important role in the demagnetization process in ferromagnets, can result from the interplay of precessional spin dynamics and spin-independent electron-phonon interaction. Further, we uncover a demagnetization regime for which the conventional Elliott-Yafet mechanism fails.

Before we discuss our approach in detail we draw attention to differences from other Elliott-Yafet-like treatments. As we neglect the small explicitly spin-dependent electronic interaction with phonons and as the longitudinal acoustic phonons, which are most important for electron-phonon scattering, do not carry angular momentum, the phonons do not take away spin in a scattering transition as envisaged in Ref.~\onlinecite{Koopmans:NatMat2010}. Thus, it is the spin-orbit coupling in the equilibrium electronic and ionic configuration (i.e., the lattice), which acts both as a spin sink and a spin source. Our approach is also fundamentally different from a recent study employing spin coherences for demagnetization dynamics~\cite{weng2017}, which inconsistently combines pure spin states with explicitly spin-dependent electron-phonon interaction matrix elements.

\section{Model} 

We employ a ferromagnetic Rashba model because it leads to simple analytical expression for the electronic single-particle states  including both spin-orbit coupling and a Stoner mean-field splitting. Compared to an \emph{ab-initio approach}, our model is much simpler and works, for numerical simplicity, with a two-dimensional $k$ space, but, in principle, is not restricted to this particular model. We use the following $2\times 2$ effective, i.e., $\bvec{k}$-dependent, hamiltonian~\cite{Fabian:2007vz}
\begin{equation}
 	\hat{H}_{\mathrm{e}}(\bvec{k})= \frac{\hbar^2 k^2}{2m_{\text{eff}}}\mathbb{1}
 		+\alpha (k_x \hat{\sigma}_y - k_y \hat{\sigma}_x) -\Delta \hat{\sigma}_z,
 			\label{eq:hamiltonian}
\end{equation} 
which determines the Bloch $u$-functions at finite $\bvec{k}$ in a two-band model. Here, the first term is a spin-diagonal kinetic contribution with effective mass~$m_{\text{eff}}$, the second is a Bychkov-Rashba spin-orbit term  with Rashba parameter $\alpha$ and the last term is the mean-field exchange splitting. The Pauli matrices are denoted by $\hat{\sigma}_{\alpha}$, with $\alpha =x$, $y$, $z$. The single-particle states of this model at each $k$ point are 2-spinors, which we denote by $|\bvec{k},\pm\rangle$ with two-dimensional $\bvec{k}=(k_x ,k_y)$. These states are used to define the reduced density matrix by $\rho_{\bvec{k}}^{\nu\nu'} = \langle c_{\bvec{k}\nu}^{\dagger} c_{\bvec{k}\nu'} \rangle$, where $c_{\bvec{k}\nu}^{\dagger}$ and $ c_{\bvec{k}\nu}$, respectively, create and annihilate an electron in the single-particle state $|\bvec{k},\nu\rangle$. The ensemble average is denoted by $\langle \cdots\rangle$. The density matrix contains the distribution functions	$n_{\bvec{k}\nu} = \rho_{\bvec{k}}^{\nu\nu}$ and the coherence $\rho_{\bvec{k}}^{+-}$. As the $|\bvec{k},\pm\rangle$ states are non-pure spin states with a $\bvec{k}$ dependent spin mixing, one can compute the $\bvec{k}$ dependent expectation values 
\begin{equation}
\langle \hat{\sigma}_{\alpha}\rangle_{\bvec{k}}=\sum_{\nu \nu'}\langle \bvec{k},\nu|\hat{\sigma }_{\alpha}|\bvec{k},\nu'\rangle \rho^{\nu\nu'}_{\bvec{k}}
	\label{eq:k-spin}
\end{equation}
The ensemble spin expectation value is determined by $S_z = (\hbar/2)\sum_{\bvec{k}}\langle \hat{\sigma}_{z} \rangle_{\bvec{k}} \equiv (\hbar/2)\langle\hat{\sigma}_z\rangle $.

The ferromagnetic character of the model comes from a Stoner model, for which we assume an effective Coulomb energy~$U_{\text{eff}}$. It determines the mean field contribution in~\eqref{eq:hamiltonian} via $\Delta =(2/3)U_{\text{eff}}\,\langle \hat{\sigma}_{z}\rangle$. 
A larger $U_{\text{eff}}$ leads to more robust ferromagnetism with a larger exchange splitting. The eigenenergies  $\epsilon_{\bvec{k}}^{\pm} = \frac{\hbar^2}{2 m_{\text{eff}}}k^2 \pm \sqrt{\Delta^{2} + \alpha^{2}k^{2}}$ are isotropic and the eigenstates take the form
\begin{equation}
	\left| \bvec{k}, \pm \right\rangle = \frac{1}{N_\pm} 
	\begin{pmatrix} - i \alpha k e^{-i \varphi_{\bvec{k}}} \\ 
	\Delta \pm \sqrt{\Delta^2 + \alpha^2 k^2}	
	\end{pmatrix}
\label{eq:EigenZstde}
\end{equation}
with $k = | \bvec{k} |$, polar angle $\varphi_{\bvec{k}}$, and normalization factor $N_\pm = \big[\alpha^2 k^2 + \big(\Delta \pm \sqrt{\Delta^2 + \alpha^2 k^2} \big)^2\big]^{1/2}$. The effective hamiltonian is formally identical to that of a spin in an external magnetic field. The Rashba contribution points in the direction $(-k_y\ k_x\ 0)^{\text{T}}$ in the $x$-$y$ plane and the Stoner mean field in the $z$ direction. The sum of the two contributions gives the $\bvec{k}$ dependent effective magnetic field. Our choice of parameters is such that the effective field is dominated by the mean-field exchange and the Rashba contribution only adds a small deviation from the~$z$ direction. The resulting band and spin structure is shown in Fig.~\ref{fig:Rashba-bands} for parameters $U_{\text{eff}}=720$\,meV and $\alpha = 30$\,meV\,nm$^{-1}$, which will be used in the numerical calculations below.

In equilibrium, we compute the single-particle states/energies together with the equilibrium density matrix self consistently for a given temperature $T_{\mathrm{eq}}$ and density $n_{\mathrm{e}}$. To this end, we assume that the equilibrium density matrix is diagonal with $n^{(\text{eq})}_{\bvec{k}\pm}=f(\epsilon_{\bvec{k}}^{\pm}-\mu)$, and we restrict the self-consistent calculation to the $z$-direction as preferred spin orientation, and thus obtain the equilibrium chemical potential~$\mu^{\text{eq}}$. With our choice of $U_{\text{eff}}$ the $|\bvec{k}, +\rangle$ states are mainly spin-up and the $|\bvec{k}, -\rangle$ are mainly spin-down, as shown in Fig.~\ref{fig:Rashba-bands}. The same parameters are used for all calculations in the paper. Because of the parabolic band structure, the mean-field equilibrium realizes a weak ferromagnet.

The time-development of the electronic spin-density matrix due to scattering with phonons is described in Markov approximation by the dynamical equation
\begin{widetext}
\begin{equation}
	\begin{split}
			&-i\hbar \frac{\partial}{\partial t} \rho_{\bvec{k}}^{\nu \nu'} = ( \epsilon_{\bvec{k}}^{\nu} - \epsilon_{\bvec{k}}^{\nu'} ) \rho_{\bvec{k}}^{\nu\nu'} \\	
			&+ \sum_{\bvec{k}_1 \nu_1} \Big\{ g_{\bvec{k}_1 \nu_1 , \bvec{k} \nu} \Big[ \sum_{\nu_2 \nu_3} g_{\bvec{k}\nu_2, \bvec{k}_1\nu_3} 
			\Big( \frac{( 1+ N_{\bvec{k}_1 -\bvec{k}}) ( \delta_{\nu_2 \nu'} - \rho_{\bvec{k}}^{\nu_2 \nu'}) \rho_{\bvec{k}_1}^{\nu_1\nu_3}  - N_{\bvec{k}_1-\bvec{k}} \left( \delta_{\nu_1 \nu_3} -\rho_{\bvec{k}_1}^{\nu_1\nu_3} \right) \rho_{\bvec{k}}^{\nu_2 \nu'} }{\epsilon_{\bvec{k}_1}^{\nu_1} - \epsilon_{\bvec{k}}^{\nu'} - \hbar\omega_{\bvec{k}_1-\bvec{k}} + i\hbar\Gamma} \\
			&\hspace{2.5cm}- \frac{( 1+ N_{\bvec{k} -\bvec{k}_1}) ( \delta_{\nu_1 \nu_3} - \rho_{\bvec{k}_1}^{\nu_1 \nu_3}) \rho_{\bvec{k}}^{\nu_2 \nu'}  - N_{\bvec{k}-\bvec{k}_1} ( \delta_{\nu_2 \nu'} -\rho_{\bvec{k}}^{\nu_2 \nu'}) \rho_{\bvec{k}_1}^{\nu_1 \nu_3} }{\epsilon_{\bvec{k}_1}^{\nu_1} - \epsilon_{\bvec{k}}^{\nu'} + \hbar\omega_{\bvec{k}-\bvec{k}_1} + i\hbar\Gamma} \Big) \Big] 
		- \big( \nu \leftrightarrow \nu' \big)^{*} \Big\}
	\end{split}	
		\label{eq:Rho-dynamics}
		\end{equation}
\end{widetext}
for the reduced density matrix. This is a standard expression that is derived, e.g., in Ref.~\onlinecite{Baral:2016cr}, under the assumption that the phonons are in equilibrium and are described by a Bose-Einstein distribution~$N_{\bvec{q}}$ for a given phonon wave vector $\bvec{q}$. An accurate calculation of the electron-phonon matrix element~$g_{\bvec{k} \nu , \bvec{k}' \nu'}$ for a ferromagnetic metal can be done \emph{ab initio}~\cite{Essert.2011,Carva:2011dp}, but for the simple model considered in this paper we make the simplifying assumption~\cite{Baral:2016cr} that the matrix element can be related to a deformation potential constant~$D$ according to $g_{\bvec{k} \nu , \bvec{k}' \nu'} =  \sqrt{\hbar/2M_{\text{ion}}c} D \sqrt{|\bvec{k}'-\bvec{k}|} \; \langle \bvec{k},\nu| \bvec{k}', \nu'\rangle $.  This is for the interaction with acoustic phonons with a linear dispersion $\hbar\omega_{\bvec{q}} = \hbar c q$, where $c$ is the sound velocity. We choose the $c$ and $M_{\text{ion}}$ values for iron. We assume $D=3.2$\,eV at first and study its influence on the dynamics below. The electrostatic electron phonon interaction is obviously spin \emph{independent}, but there are electron-phonon ``spin-flip'' matrix elements $g_{\bvec{k} + , \bvec{k}' -}$ because the spin-orbit coupling gives rise to nonvanishing overlaps $\langle \bvec{k},+|\bvec{k}',-\rangle\neq0$. 

The effective hamiltonian~\eqref{eq:hamiltonian} does not commute with the spin operator~$\hat{\sigma}_{z}$ because of the spin-orbit coupling, and enters Eq.~\eqref{eq:Rho-dynamics} directly via the spin splitting between the electronic energies~$\epsilon_{\bvec{k}}^{+}-\epsilon_{\bvec{k}}^{-}$. This spin splitting leads to a contribution to the equation of motion~\eqref{eq:Rho-dynamics} for the coherence $\rho^{+-}$ that describes the precession of the spin expectation value~\eqref{eq:k-spin} around the $\bvec{k}$-dependent effective internal field. Even though the precessional contribution is always present, it only leads to an oscillatory dynamics if it is not counteracted by the scattering term in~\eqref{eq:Rho-dynamics} .

\section{Results} 

For the calculation of the spin and charge dynamics we do not attempt to model the details of ultrashort-optical-pulse excitation here, but choose the following simple initial conditions. We take the self-consistently determined equilibrium states/energies and change the electronic distributions in the equilibrium density matrix instantaneously to an elevated temperature $T_{\text{e}} > T_{\text{eq}}$ while keeping the electron density fixed, which results in a small change of the ensemble spin. In the following we always assume~$T_{\text{eq}}=70$\,K.

\begin{figure}		
	\includegraphics[scale=0.4]{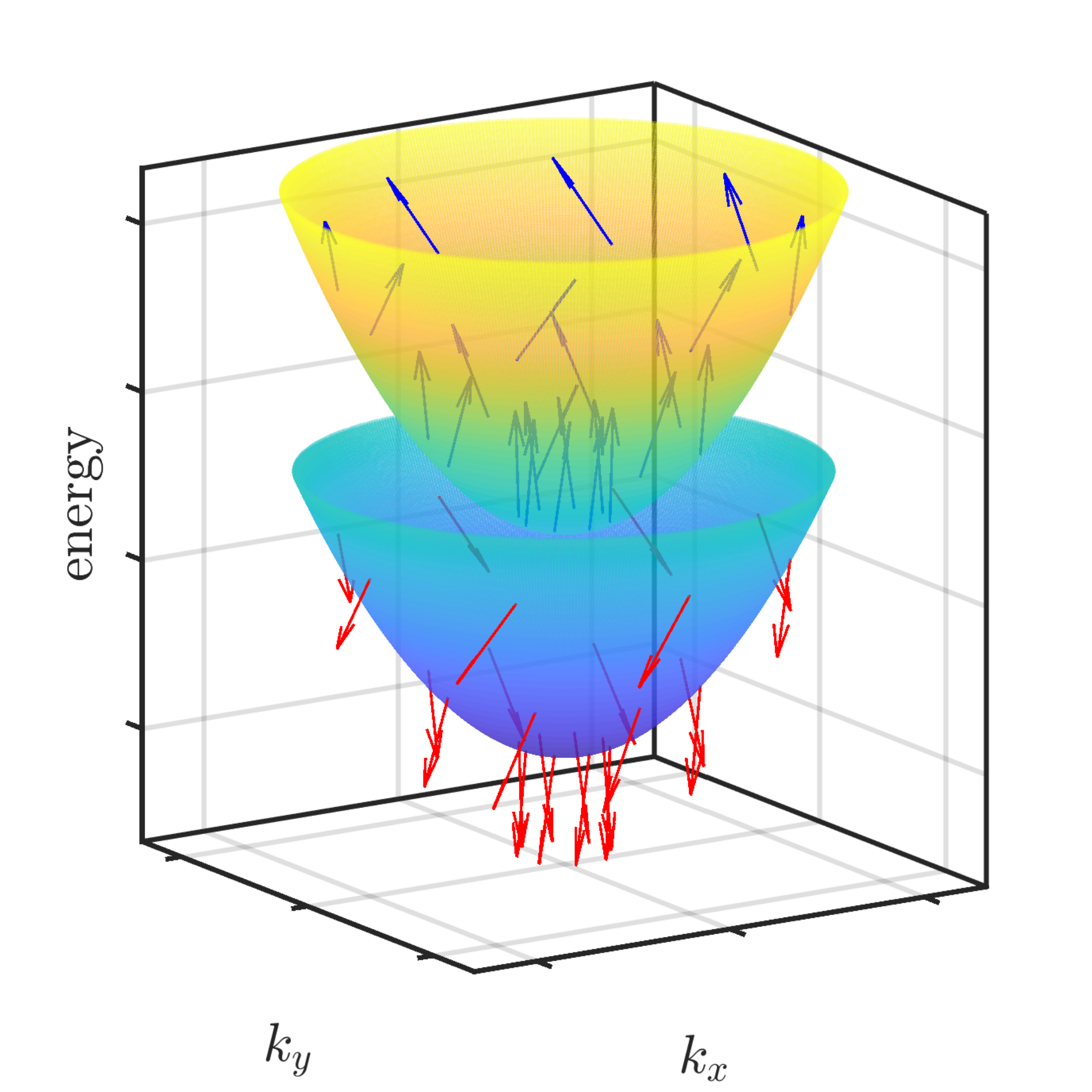}		
	\caption{Band structure of the ferromagnetic Rashba model for an equilibrium temperature $T_{\mathrm{eq}}=70\,\mathrm{K}$, Stoner parameter $U_{\text{eff}}=720$\,meV, and Rashba parameter $\alpha = 30$\,meV\,nm$^{-1}$. The expectation values $\langle \bvec{k}\nu|\vec{\hat{\sigma}}|\bvec{k}\nu\rangle$ at some $\bvec{k}$ points are indicated by arrows and the energy by a color code: low (dark blue) to high (light yellow). 
		\label{fig:Rashba-bands}}
\end{figure}

\begin{figure}[t]
	\includegraphics[scale=0.8]{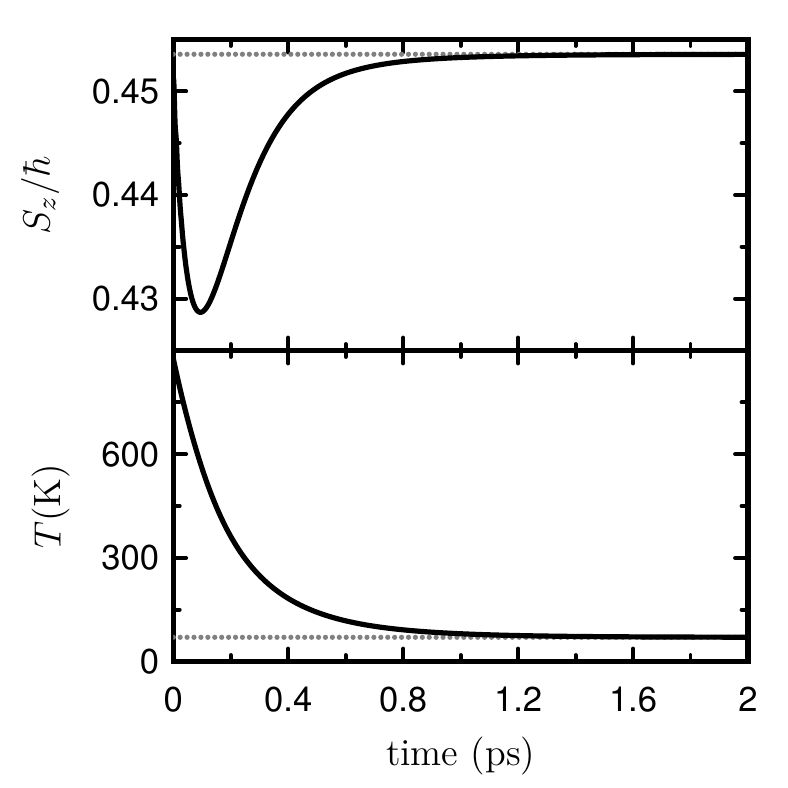}
	\caption{Dynamics of the $z$ component of the ensemble spin $S_z$, and the electronic temperature $T$ of the full EOM \eqref{eq:Rho-dynamics} for parameters as in Fig.~\ref{fig:Rashba-bands} and an excitation temperature $T_{\mathrm{ex}}=1000\, \mathrm{K}$. Equilibrium values are indicated by dotted horizontal lines.}
	\label{fig:dynamics}
\end{figure}

Figure~\ref{fig:dynamics} shows the ensemble spin and charge dynamics resulting from Eq.~\eqref{eq:Rho-dynamics} for an excited electronic kinetic energy (temperature) of $T_{\text{e}}=1000$\,K. The numerical evaluation of~\eqref{eq:Rho-dynamics} assumes an infinitesimal broadening~$\hbar\Gamma\to 0$. 
We characterize the charge dynamics by fitting the non-equilibrium distributions $n_{\bvec{k},\pm}$ with Fermi-Dirac functions where the fit parameters are a common temperature and different chemical potentials for the two bands. Fig.~\ref{fig:dynamics}(a) shows a demagnetization in about 100\,fs, followed by a slower remagnetization. The microscopic charge scattering dynamics is reflected in the change of the effective temperature (determined for all electrons in the system), shown in Fig.~\ref{fig:dynamics}(b). 

An important result of our calculation is that, microscopically, demagnetization dynamics due to electron-phonon interaction in ferromagnets occurs via the \emph{interplay} between scattering and $k$-dependent internal magnetic fields. This can be demonstrated as follows: Analytically one can show that the scattering contribution in~\eqref{eq:Rho-dynamics} \emph{alone} does not lead to any spin/magnetization dynamics, i.e., for the ensemble spin~$S_z$ we find $dS_z/dt = 0$~\footnote{A numerical solution of~\eqref{eq:Rho-dynamics} for that case also shows that the temperature does not completely return to its equilibrium value at later times.}, if we leave out the coherent precession~\cite{Wu:2000epjb}. We have checked this numerically by switching off the precessional term, i.e., the first term on the RHS of~\eqref{eq:Rho-dynamics}, which results in a similar charge scattering  dynamics after excitation, but no magnetization dynamics at all. Further, the precessional term alone, can, in principle, change the magnetization by dephasing as it describes precession around the effective magnetic fields, but results only in a very small magnetization change that is four orders of magnitude smaller than what is shown in Fig.~\ref{fig:dynamics}(a). It therefore has to be the combination of both contributions. In semiconductors, a similar spin relaxation mechanism was first identified by Wu and coworkers~\cite{Wu:2000epjb,Wu:2010jf}. 

\begin{figure}
	\includegraphics[width=0.4\textwidth]{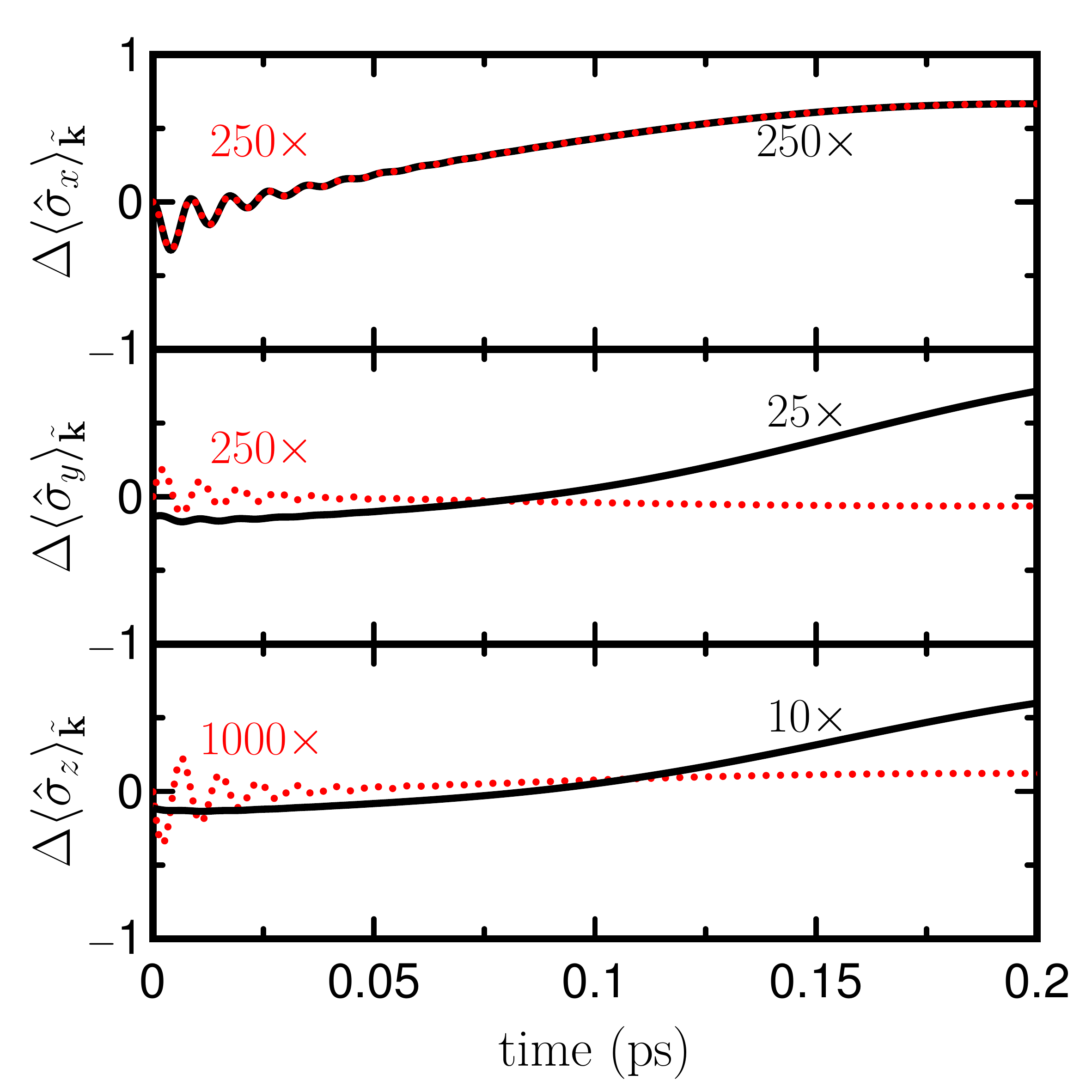}
	\caption{Computed spin change $\Delta \langle \hat{\sigma}_{\alpha}\rangle_{\tilde{\bvec{k}}}$ at $\tilde{\bvec{k}}$ (solid line). At this point in $k$-space the effective internal magnetic field points in the $y$-$z$ direction. Dotted red line: contribution to~\eqref{eq:k-spin} only from the coherences~$\rho^{+-}_{\bvec{k}}$. Note that in the lower two graphs the different curves are multiplied by different factors as indicated, whereas both curves are identical in the top graph. The parameters are the same as in Fig.~\ref{fig:dynamics}.
	\label{fig:precession}}
\end{figure}

We elucidate the interplay of scattering and precession in Fig.~\ref{fig:precession} by investigating the (spin) expectation value~\eqref{eq:k-spin} for a $k$ space cell at a particular point $\tilde{\bvec{k}}$, which lies just above the Fermi wave vector of the ``$-$'' band in positive $k_x$ direction. In equilibrium, the expectation value~\eqref{eq:k-spin} points in the \emph{direction} of the local effective field, which is also the \emph{direction} indicated by the arrows in Fig.~\ref{fig:Rashba-bands} at each $\bvec{k}$ point. At $\tilde{\bvec{k}}$, in particular, this direction lies in $y$-$z$ plane. We plot the change  $\Delta \langle \hat{\sigma}_{\alpha}\rangle_{\tilde{\bvec{k}}}(t)=\langle \hat{\sigma}_{\alpha}\rangle_{\tilde{\bvec{k}}}(t) -\langle \hat{\sigma}_{\alpha}\rangle^{(\text{eq})}_{\tilde{\bvec{k}}} $ with respect to the equilibrium expectation value. In addition, we plot the contribution to~$\Delta \langle \hat{\sigma}_{\alpha}\rangle_{\tilde{\bvec{k}}}(t)$ arising from the spin coherences, i.e., the $\nu\neq\nu'$ terms in~\eqref{eq:k-spin}, only, and we discuss these first. For our excitation conditions, the coherences are initially zero and the scattering contribution to Eq.~\eqref{eq:Rho-dynamics} is needed to get the coherences started. However, their dynamics is also influenced by the precessional contribution in~\eqref{eq:Rho-dynamics} due to  the effective magnetic field. Note that the precessional contribution is always there (with a period of about 7\,fs mainly due to the Stoner contribution to the spin splitting), but it results in oscillatory motion of the coherences with that period only during the first 50\,fs. The first 50\,fs, during which also the main demagnetization happens, is the time scale on which the system reaches a quasi-equilibrium in the sense that electrons in both bands can be described by the same effective temperature (but the respective densities are different from their equilibrium values). After that, the scattering suppresses the oscillatory motion of the coherences, but the spin dynamics stay noncollinear, as can be seen from the $x$ component at $\tilde{\bvec{k}}$. It is nonzero not only on the time scale of Fig.~\ref{fig:precession}, but also during the remagnetization, and its existence demonstrates that the spin expectation value does not point into the direction of the effective magnetic field at $\tilde{\bvec{k}}$. Note that, while the spin precession contribution from the coherences is important for the spin dynamics, it is essentially invisible on the scale of $\Delta \langle \hat{\sigma}_{y}\rangle_{\tilde{\bvec{k}}}(t)$ and $\Delta \langle \hat{\sigma}_{z}\rangle_{\tilde{\bvec{k}}}(t)$, which are dominated by the distributions~$n_{\pm\bvec{k}}$.

\begin{figure}[t]	 		
	\includegraphics[width=0.4\textwidth]{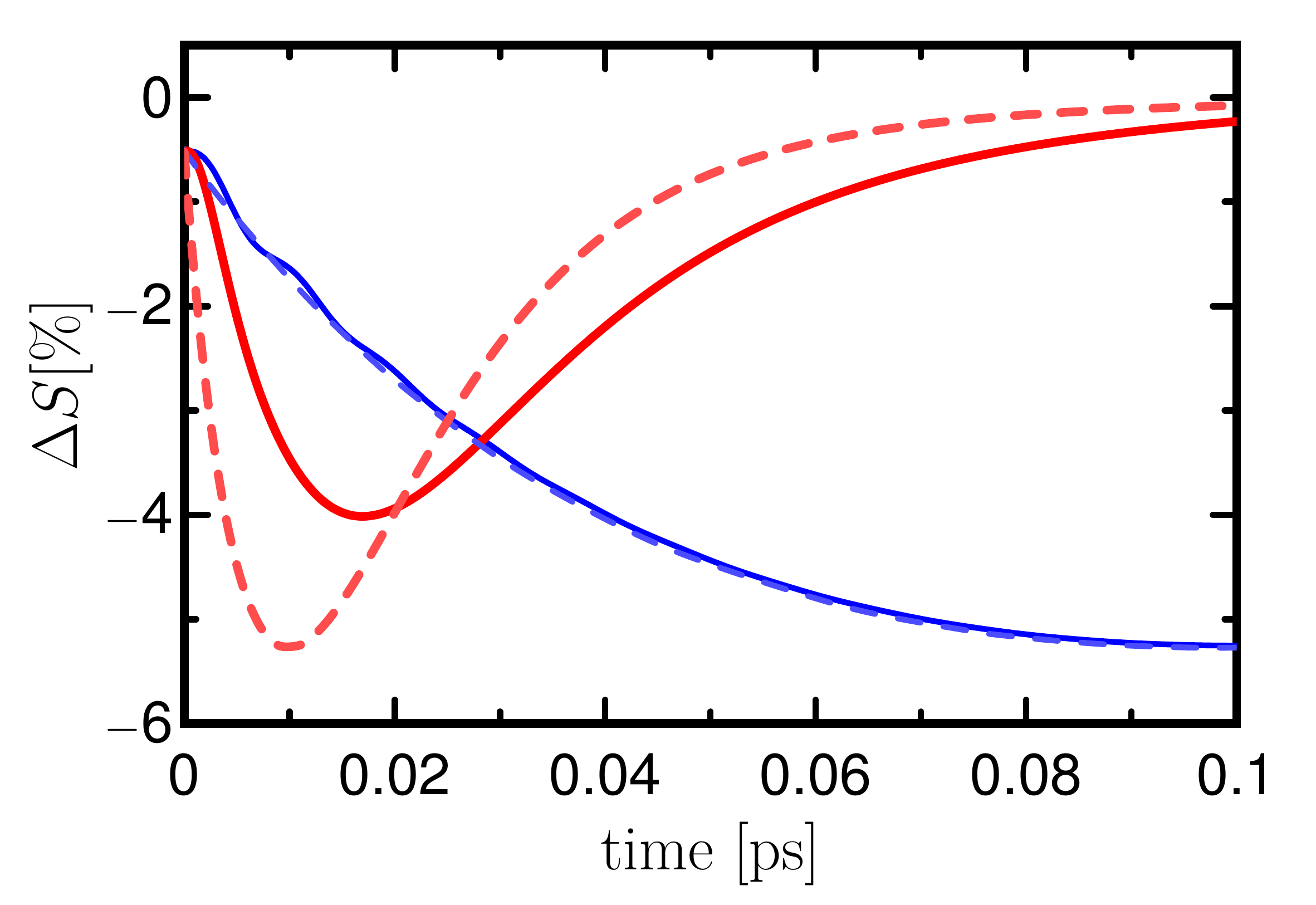}
	\caption{Computed normalized spin dynamics $\Delta S= (S_z - S_z^{(\text{eq})})/S_z^{(\text{eq})}$ calculated from Eq.~\eqref{eq:Rho-dynamics} (solid lines) and without coherences, i.e., including only electronic distributions (dashed lines). The thin blue curves are the same as in Fig.~\ref{fig:dynamics}, for the thick red curves we enlarged the deformation potential $D^2\to 10\,D^2$. All other parameters as in Fig.~\ref{fig:dynamics}.
		\label{fig:dynamics-Ds}}
\end{figure}

Finally, we would like to explore the connection between spin precession and spin-flip scattering (Elliott-Yafet mechanism) in Fig.~\ref{fig:dynamics-Ds} for different electron-phonon coupling strengths. To do this we compare the computed dynamics using the full Eq.~\eqref{eq:Rho-dynamics} for the reduced density matrix~$\rho$ to that neglecting the coherences $\rho^{+-}_{\bvec{k}}$ (and therefore also the precession) in~Eq.~\ref{eq:Rho-dynamics}. In the latter case, we keep only the diagonal parts of the density matrix, i.e., the distribution functions~$n_{\bvec{k}\pm}$ whose dynamics are governed by conventional Boltzmann scattering integrals~\cite{Krauss:2009gc,Essert.2011,Mueller.2013}. Note that without the coherences, only components of the spin expectation value in the direction of the local effective magnetic field are included and the ensemble spin is given by $S_z=(\hbar/2)\sum_{\bvec{k}\nu}\langle \bvec{k}, \nu|\hat{\sigma}_z|\bvec{k}, \nu\rangle n_{\bvec{k}\nu}$.

Fig.~\ref{fig:dynamics-Ds} shows that, for the set of parameters used so far, the calculations with and without coherences lead to almost identical spin dynamics. The reason for this similarity lies in the precessional contribution to the equation of motion~\eqref{eq:Rho-dynamics} for the coherences discussed in connection with Fig.~\ref{fig:precession}, even though the precessional dynamics is only visible for short times due to the influence of the scattering. For the electron-phonon coupling strength used so far, the precession period period of 7\,fs is short compared to the energy relaxation time of 120\,fs, so that, at each $k$ point, the part of the spin expectation value deviating from the (local) magnetic field direction averages out. It therefore looks like as if the spin had instantaneously changed (i.e., ``flipped'') into its projection on the direction of the effective magnetic field, as assumed in the conventional Elliott-Yafet mechanism. Even though the spin dynamics are almost identical, the underlying microscopic dynamics are still different. In particular, the distribution functions $n^{\pm}_{\bvec{k}}$ in both calculation differ by a small amount.

We next turn to the spin dynamics when the deformation potential is increased $D^2\to10\,D^2$. In this case, calculation with coherences leads to a reduced demagnetization with a smaller minimum at slightly longer times, as compared to the calculation neglecting coherences. As the energy relaxation time is shortened by an order of magnitude to 12\,fs, the precessional contribution to~\eqref{eq:Rho-dynamics} no longer just leads to an effective projection of $\langle \bvec{s}\rangle_{\bvec{k}}$ on the direction of the effective field. For this case, the conventional Elliott-Yafet-type picture with spin flips fails, and cannot be replaced by a simpler description. Only if the electron-phonon coupling strength were increased more so that the scattering becomes faster than the precession, one enters a collision dominated regime that could be described by the classical Dyakonov-Perel type picture with motional narrowing. 
  
Our results show that calculations of the spin/magnetization dynamics using the conventional Elliott-Yafet mechanism~\cite{Carva:2011dp,Essert.2011} can only \emph{overestimate} the demagnetization due to electron-phonon scattering. Thus the conclusion of these earlier papers that this process is not efficient enough to explain the experimentally observed ultrafast magnetization dynamics remain unchanged~\footnote{If one includes a time-dependent (mean-field) exchange splitting~\cite{Mueller.2013}, the effect of spin-flip scattering of electrons with phonons and other electrons can be amplified to reach magnitudes observed in experiments.}. However, even if the calculations with and without coherences yield similar spin dynamics, the presence of spin coherences gives a slightly different picture of how the lattice acts a spin sink and spin source via the precession around internal effective magnetic fields. This understanding is in line with the variety of spin-orbit induced transport effects~\cite{Sinova:2004prl}. 

\section{Conclusion}
 We investigated ultrafast magnetization dynamics in a ferromagnetic model system including spin-orbit coupling and electron-phonon scattering. By computing the reduced spin density matrix for itinerant electrons we showed how the magnetization change occurs due to the interplay of spin precession around internal effective magnetic fields and spin-independent scattering. If the precession period around the exchange field is short compared to typical scattering time, the precessional contributions are effectively averaged out and one obtains good agreement with the magnetization computed using spin-flip transition rates, as assumed in the conventional Elliott-Yafet mechanism. For shorter scattering times, we find a magnetization dynamics that is slower and less pronounced compared to the conventional Elliott-Yafet mechanism.

\begin{acknowledgements} 
This work was supported by the DFG through the SFB/TRR 173 ``Spin+X'' (Project A8).
\end{acknowledgements}

\bibliographystyle{apsrev4-1}
\bibliography{EY,demagnetization,switching}

\end{document}